\documentclass[twocolumn]{jpsj2} %% two-column layout
\title{Multiband Superconductivity in Filled-Skutterudite Compounds (Pr$_{1-x}$La$_{x}$)Os$_{4}$Sb$_{12}$: An Sb Nuclear-Quadruple-Resonance (NQR) Study}

\author{
Mamoru \textsc{Yogi}\thanks{Present address: Department of Physics and Earth Sciences, Faculty of Science, University of the Ryukyus, Okinawa 903-0213}\thanks{E-mail address: myogi@sci.u-ryukyu.ac.jp},
Takayuki \textsc{Nagai},
Yojyu \textsc{Imamura},
Hidekazu \textsc{Mukuda},
Yoshio \textsc{Kitaoka}\thanks{E-mail address: kitaoka@mp.es.osaka-u.ac.jp},
Daisuke \textsc{Kikuchi}$^{1}$,
Hitoshi \textsc{Sugawara}$^{2}$,
Yuji \textsc{Aoki}$^{1}$,
Hideyuki \textsc{Sato}$^{1}$,
and Hisatomo \textsc{Harima}$^{3}$
}

\inst{
Department of Materials Engineering Science, Graduate School of Engineering Science, Osaka University, Osaka 560-8531\\
$^{1}$Department of Physics, Tokyo Metropolitan University, Tokyo 192-0397\\
$^{2}$Department of Mathematical and Natural Sciences, Faculty of the Integrated Arts and Sciences, The University of Tokushima, Tokushima 770-8502\\
$^{3}$Department of Physics, Faculty of Science, Kobe University, Kobe 657-8501
}

\abst{
We report on the systematic evolution of normal-state properties and superconducting characteristics in filled-skutterudite compounds (Pr$_{1-x}$La$_{x}$)Os$_{4}$Sb$_{12}$ determined using Sb nuclear-quadrupole-resonance (NQR) experiments.
The Sb-NQR spectra in these compounds have split into two sets, arising from different Sb$_{12}$ cages containing either Pr or La, which enables us to measure two kinds of nuclear spin-lattice relaxation time $T_{1}^{Pr}$ and $T_{1}^{La}$.
In the normal state, the temperature ($T$) dependence of $1/T^{Pr}_1T$ showed almost the same behavior as that for pure PrOs$_{4}$Sb$_{12}$ regardless of the increase in La content.
In contrast, $1/T^{La}_1T$ markedly decreases with increasing La concentration.
These results show that $4f^{2}$-derived magnetic fluctuations are almost localized at the Pr site.
In the superconducting state for Pr-rich compounds of $x=0.05$ and 0.2, $1/T_{1}^{Pr}$ exponentially decreases down to $T=0.7$ K with no coherence peak below $T_{c}$ as well as in PrOs$_{4}$Sb$_{12}$.
A remarkable finding is that the residual density of states (RDOS) at the Fermi level below $T_c$ is induced by La substitution for Pr.
The impurity effect, usually observed in unconventional superconductors with a line-node gap, may not be the origin of the RDOS induced by the La substitution, since RDOS does not increase and $T_{c}$ does not decrease with increasing La content.
RDOS is more naturally explained if a small part ($\sim$ 5.5\%) of the total Fermi surface (FS) becomes gapless for $x=0.05$ and 0.2.
These results are proposed to be understood in terms of a multiband-superconductivity (MBSC) model that assumes a full gap for part of the FS and the presence of point nodes for a small $4f^2$-derived FS inherent in PrOs$_{4}$Sb$_{12}$.
The former could be relevant with FS existing in LaOs$_{4}$Sb$_{12}$ and with the anisotropic gap with point nodes being markedly suppressed by either applying a magnetic field or substituting La for Pr.
For La-rich compounds of $x=0.8$ and 1, on the other hand, $1/T^{La}_1$ exhibits a coherence peak and the  nodeless energy gap characteristic for weak-coupling BCS $s$-wave superconductors.
With increasing Pr content, $T_c$ increases and the energy gap increases from $2\Delta_{0}/k_{\rm B}T_{c}= 3.45$ for pure La compounds up to $2\Delta_{0}/k_{\rm B}T_{c}=4.2$ and 5.2 for the 60\% Pr and 80\% Pr compounds, respectively.
The Pr substitution for La enhances the pairing interaction and induces an anisotropy in the energy-gap structure.
The novel strong-coupling superconductivity in PrOs$_{4}$Sb$_{12}$ is inferred to be mediated by the local interaction between  $4f^{2}$-derived crystal-electric-field states with the electric quadrupole degree of freedom and conduction electrons.
This coupling causes a mass enhancement of quasi-particles for a part of FS and induces a small FS, which is responsible for point nodes in the superconducting gap function.
Note that the small FS does not play any primary role for the strong-coupling superconductivity in PrOs$_{4}$Sb$_{12}$.
}

\kword{filled skutterudite, heavy fermion, superconductivity, impurity effect, NQR, PrOs$_{4}$Sb$_{12}$}

\begin{document}
\maketitle
\newpage

\section{Introduction}
A series of filled-skutterudite compounds have attracted much interest because of their unique properties.
For instance, Pr-based compounds show diverse physical properties such as metal-insulator transition, antiferroquadrupolar order, and heavy-fermion (HF) superconductivity.\cite{Sekine, Iwasa, Sato, Tayama, Ishida, Takeda, Yogi}
Among them, PrOs$_{4}$Sb$_{12}$ is the first example of a Pr-based HF superconductor with $T_c=1.85$ K.\cite{Bauer1}
The superconducting (SC) characteristics of PrOs$_{4}$Sb$_{12}$ differ significantly from those of Ce- and U-based unconventional HF SC compounds.
The Sb nuclear-quadrupole-resonance (NQR) measurements showed an exponential decrease in the nuclear spin-lattice relaxation rate $1/T_{1}$ with no coherence peak at low temperatures associated with a well-developed gap.\cite{Kotegawa} 
The London-penetration depth probed by muon-spin relaxation was reported to decrease exponentially with decreasing temperature.\cite{MacLaughlin}
Scanning tunneling spectroscopy also showed that an SC gap exists over a large part of the Fermi surface (FS).\cite{Suderow}
The deviations from isotropic BCS $s$-wave behavior were however discussed in terms of a finite distribution of SC gaps.

On the other hand, some other experiments revealed point nodes in the gap function; a London-penetration-depth study showed a quadratic power-law behavior of $\lambda$; thermal conductivity $\kappa$ measurement in a rotated magnetic field revealed the presence of two distinct SC phases with twofold and fourfold symmetries with point nodes.\cite{Chia, Izawa}
Furthermore, another SC state below $T\approx 0.6$ K ($T/T_{c}\approx 0.3$) was suggested from pronounced enhancements in lower critical field $H_{c1}(T)$ and critical current $I_{c}(T)$.\cite{Cichorek}
In addition, the recent thermal transport measurement under a magnetic field pointed to a possibility of a multiband superconductivity (MBSC) from a unique magnetic field dependence of $\kappa$ similar to MgB$_{2}$.\cite{Seyfarth}
From these results, an order-parameter symmetry and a relevant energy-gap structure for PrOs$_{4}$Sb$_{12}$ are not fully understood yet.
Apparently, these results differ from those of most other HF superconductors with a line node in the gap function.\cite{Tou, Ishida_CeCuSi, Zheng, YuK}

In contrast to Ce-based HF superconductors with the line-node gap, the crystal electric field (CEF) levels in PrOs$_{4}$Sb$_{12}$ have a singlet-$\Gamma_{1}$ ground state accompanying the very low-lying first-excited $\Gamma_{4}^{(2)}$ state,\cite{Kohgi, Goremychkin} with a quadrupole degree of freedom.
Therefore, the HF-like behavior exhibited by PrOs$_4$Sb$_{12}$ is considered due to a very small energy splitting $\Delta_{\rm CEF}/k_{\rm B} \sim 8$ K between $\Gamma_{1}$ and $\Gamma_{4}^{(2)}$.
This is because PrRu$_{4}$Sb$_{12}$ is the conventional $s$-wave superconductor when the energy splitting $\Delta_{\rm CEF}/k_{\rm B} \sim 70$ K is one order of magnitude larger than that for PrOs$_{4}$Sb$_{12}$.\cite{Takeda, Yogi}
Thus, in Pr-based compounds, the $4f^2$-derived CEF effect plays an important role in their rich physical phenomena.
Various experiments were reported for a series of Pr(Os$_{1-x}$Ru$_{x}$)$_{4}$Sb$_{12}$ compounds,\cite{Frederick1, Frederick2, Nishiyama} where SC takes place for all Ru concentrations $x$.
A minimum of $T_c$ in these compounds is supposed to arise from the competition between the HF superconductivity of PrOs$_{4}$Sb$_{12}$ and the BCS superconductivity of PrRu$_{4}$Sb$_{12}$.
Sb-NQR $T_1$ measurements provided evidence of the existence of nodes in the SC gap for Pr(Os$_{1-x}$Ru$_{x}$)$_{4}$Sb$_{12}$.\cite{Nishiyama}
In contrast, a series of compounds (Pr$_{1-x}$La$_{x}$)Os$_{4}$Sb$_{12}$ provide an interesting opportunity for obtaining insight into SC characteristics by abolishing the $4f^2$ derived coherence effect in pure compounds.
This is because La substitution for Pr  does not change $\Delta_{\rm CEF}$ significantly, but disturbs the periodicity of Pr-$4f^2$-derived periodic lattice.\cite{Rotundu}

The unique crystal structure of the filled-skutterudite (space group \textit{Im\={3}}) is that the Pr ion is located at the center of the cage formed by neighboring 12 Sb ions.
The variation in the lattice constant for (Pr$_{1-x}$La$_{x}$)Os$_{4}$Sb$_{12}$ is very small amounting to $\sim 0.05$\% between the end components.\cite{Bauer2, Sugawara, Harima} 
LaOs$_{4}$Sb$_{12}$ shows a conventional $s$-wave superconductivity with $T_{c}=0.74$ K,\cite{Kotegawa} whereas PrOs$_{4}$Sb$_{12}$ shows a novel superconductivity that differs from the previous examples showing an unconventional superconductivity with the line-node gap.
Note that $T_{c}$ for a series of (Pr$_{1-x}$La$_{x}$)Os$_{4}$Sb$_{12}$ compounds decreases  almost linearly as  La concentration increases.\cite{Rotundu}
This result is in contrast with the well known fact that the unconventional HF superconductors undergo a marked depression of $T_{c}$ by substituting La for Ce which eliminates the $4f$-electrons derived coherence effect in pure compounds.\cite{Vorenkamp, Dalichaouch, Petrovic}
In this paper, we report on the intimate evolution of electronic and SC characteristics in a series of (Pr$_{1-x}$La$_{x}$)Os$_{4}$Sb$_{12}$ compounds through the Sb-NQR  measurements.

\section{Experimental Procedures}

\begin{figure}[tbh]
  \begin{center}
    \includegraphics[keepaspectratio=true,height=45mm]{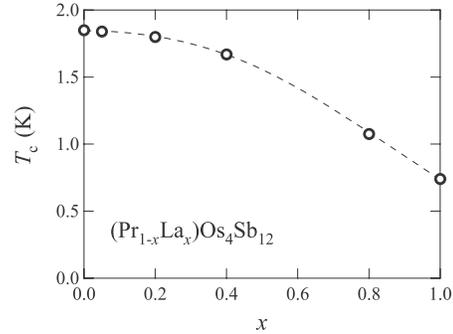}
  \end{center}
  \caption{SC transition temperature $T_{c}$ as function of La concentration $x$. The broken line is a visual guide.}
\end{figure}

Single crystals of PrOs$_{4}$Sb$_{12}$ and LaOs$_{4}$Sb$_{12}$ were prepared by the Sb-flux method.\cite{Sugawara}
The high quality of the sample was demonstrated by the observation of de Haas-van Alphen oscillations.\cite{Sugawara}
A series of compounds (Pr$_{1-x}$La$_{x}$)Os$_{4}$Sb$_{12}$ ($x = 0.05, 0.2, 0.4, 0.8$) were prepared by the same method.
The SC transition temperature $T_{c}$ shown in Fig. 1 was determined by the midpoint of resistivity drop; $T_{c}=$ 1.85, 1.84, 1.80, 1.67, 1.07 and 0.74 K for $x=0$, 0.05, 0.2, 0.4, 0.8 and 1, respectively.
For the Sb-NQR measurement, the samples were powdered to facilitate applied rf-field penetration.
The Sb-NQR measurement was performed using the conventional spin-echo method at $H=0$ and in the temperature ($T$) range of $T = 0.15-200$ K using a $^3$He-$^{4}$He-dilution refrigerator.
The NQR spectrum was obtained by integrating a spin-echo signal point by point as a function of frequency.
The nuclear spin-lattice relaxation time $T_{1}$ was obtained by the conventional saturation recovery method.

\section{Results and Analyses}

\subsection{NQR spectrum}

\begin{figure}[tbh]
  \begin{center}
    \includegraphics[keepaspectratio=true,height=75mm]{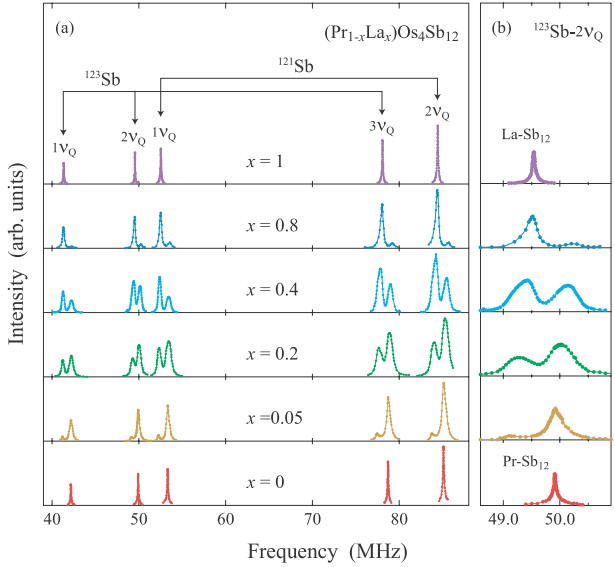}
  \end{center}
  \caption{(a) $^{121}$Sb- and $^{123}$Sb-NQR spectra for (Pr$_{1-x}$La$_{x}$)Os$_{4}$Sb$_{12}$ at 4.2 K. (b) An enlarged view of spectra at $^{123}$Sb-$2\nu_{Q}$ ($\pm5/2 \leftrightarrow \pm3/2$) transitions.}
\end{figure}

\begin{fulltable}[tb]
\caption{Experimental and calculated\cite{Harima} values of $^{121}\nu_{Q}$, $^{123}\nu_{Q}$ and $\eta$ for ReTx$_{4}$Sb$_{12}$ (Re = La, Ce and Pr; Tx = Ru and Os).}
\label{t1}
\begin{tabular}{cccccccc}
\hline
 & \multicolumn{2}{c}{$^{121}\nu_{Q}$ (MHz)} & \multicolumn{2}{c}{$^{123}\nu_{Q}$ (MHz)} & \multicolumn{2}{c}{$\eta$} \\
\cline{2-8}
material & Exp. & Calc. & Exp. & Calc. & Exp. & Calc. & Reference \\
\hline
LaRu$_{4}$Sb$_{12}$ & 41.171  & 42.118  & 24.994  & 25.171  & 0.406  & 0.3816  & \\
CeRu$_{4}$Sb$_{12}$ & 41.358  & 42.698  & 25.108  & 25.846  & 0.401  & 0.3732  & \\
PrRu$_{4}$Sb$_{12}$ & 41.516  & 42.267  & 25.204  & 25.585  & 0.402  & 0.4341  & \cite{yogiPrRuSb}\\
 &  &  &  &  &  &  & \\
LaOs$_{4}$Sb$_{12}$ & 43.777  & 43.138  & 26.576  & 26.113  & 0.450  & 0.4503  & \\
CeOs$_{4}$Sb$_{12}$ & 43.847  & 43.502  & 26.628  & 26.333  & 0.463  & 0.4523  & \cite{yogiCeOsSb}\\
PrOs$_{4}$Sb$_{12}$ & 44.175  & 43.686  & 26.817  & 26.444  & 0.459  & 0.5281  & \cite{Kotegawa}\\
\hline
\end{tabular}

\end{fulltable}

Figure 2 shows the Sb-NQR spectra of (Pr$_{1-x}$La$_{x}$)Os$_{4}$Sb$_{12}$ ($x = 0$, 0.05, 0.2, 0.4, 0.8 and 1) at 4.2 K.
Five NQR transitions are resolved for pure compounds ($x = 0$ and 1).
These five NQR transitions arise from the Sb isotopes $^{121}$Sb and $^{123}$Sb with respective natural abundances of 57.3\% and 42.7\%, the nuclear spins $I$ = 5/2 and 7/2, the nuclear quadrupole moments $^{121}Q$ and $^{123}Q$.
Thus, two and three NQR transitions arise from $^{121}$Sb and $^{123}$Sb, respectively.
The NQR Hamiltonian is described as
\begin{equation}
{\cal H}_{Q}=\frac{h\nu_{Q}}{6}[3I_{z}^{2}-I^{2}+\frac{\eta}{2}(I_{+}^{2}+I_{-}^{2})],
\end{equation}
with
\begin{equation}
\nu_{Q}\equiv \frac{3e^{2}qQ}{2I(2I-1)h}
\end{equation}
and
\begin{equation}
\eta\equiv \frac{|V_{xx}-V_{yy}|}{V_{zz}},
\end{equation}
where $I$ is the nuclear spin, $\nu_{Q}$ is the nuclear quadrupole frequency, and $\eta$ is the asymmetry parameter.
Here, $V_{xx}$, $V_{yy}$, and $V_{zz}$ are three components of the electric field gradient (EFG) tensor.
Table \ref{t1} shows a summary of the $^{121}\nu_{Q}$, $^{123}\nu_{Q}$ and  $\eta$ obtained from the NQR spectra of the filled-skutterudite antimonide compounds ReTx$_{4}$Sb$_{12}$ (Re = La, Ce, Pr; Tx = Os, Ru) together with those estimated through band calculation by means of FLAPW-LDA.\cite{Kotegawa,Harima,yogiPrRuSb,yogiCeOsSb}
The $\nu_{Q}$ and $\eta$ of PrOs$_{4}$Sb$_{12}$ obtained in this study are the same as those reported by Kotegawa \textit{et al}.\cite{Kotegawa} 
Considering the fact that all the compounds indicate the same ratio of $^{123}\nu_{Q}/^{121}\nu_{Q} = 0.607$, a nuclear quadrupole moment ratio of $^{123}Q/^{121}Q =$ 1.275 is experimentally deduced.
The calculated values are in good agreement with the experimental ones.

The Sb-NQR spectra of (Pr$_{1-x}$La$_{x}$)Os$_{4}$Sb$_{12}$ split into two sets, arising from Sb$_{12}$ cages occupied by Pr and La (hereafter denoted as Pr and La cages, respectively).
The Sb sites for Pr and La cages are denoted as Sb(1) and Sb(2), respectively.
The low-frequency and high-frequency peaks in the two sets of spectra arise from Sb(2) and Sb(1), respectively, because the intensity of Sb(2) at the low-frequency peak becomes larger than that of the high-frequency one as La concentration increases as shown in Figs. 2(a) and 2(b).
Thus site-selective NQR measurements enable us to extract the evolution of the local electronic state in Sb(1) and Sb(2).
On the other hand, since the difference in lattice constant between PrOs$_{4}$Sb$_{12}$ and PrRu$_{4}$Sb$_{12}$ is larger than that between PrOs$_{4}$Sb$_{12}$ and LaOs$_{4}$Sb$_{12}$, the NQR spectra of Pr(Os$_{1-x}$Ru$_{x}$)$_{4}$Sb$_{12}$ are broader with more complicated spectral shape than those of the present compounds.\cite{Nishiyama}
In this context, Ru substitution for Os causes a large distribution of EFGs, giving rise to some crystal defects or crystal imperfection.
The values of $^{123}\nu_{Q}$ and $\eta$ at 4.2 K for (Pr$_{1-x}$La$_{x}$)Os$_{4}$Sb$_{12}$ are summarized in table \ref{t2}.

\begin{table}[tb]
\caption{Experimental values of $^{123}\nu_{Q}$ and $\eta$ for (Pr$_{1-x}$La$_{x}$)Os$_{4}$Sb$_{12}$.}
\label{t2}
\begin{center}
\begin{tabular}{ccccc}
\hline
 & \multicolumn{2}{c}{Sb cage with Pr} & \multicolumn{2}{c}{Sb cage with La} \\
\cline{2-5}
$x$ & $^{123}\nu_{Q}$ (MHz) & $\eta$ & $^{123}\nu_{Q}$ (MHz) & $\eta$ \\
\hline
0 & 26.817 & 0.459  &  &  \\
0.05 & 26.83  & 0.459  & 26.39  & 0.453  \\
0.2 & 26.86  & 0.458  & 26.45  & 0.452  \\
0.4 & 26.93  & 0.456  & 26.51  & 0.451  \\
0.8 & 26.97  & 0.456  & 26.56  & 0.450  \\
1 &  &  & 26.576 & 0.450  \\
\hline
\end{tabular}
\end{center}
\end{table}

\subsection{Sample dependence of $1/T_{1}$ for PrOs$_{4}$Sb$_{12}$ }

First, we deal with the sample dependence of $1/T_{1}$ for PrOs$_{4}$Sb$_{12}$ which was reported to be saturated at low temperatures well below $T_c$ in our previous work.\cite{Kotegawa}
The sample in our previous work was crushed into fine powder for the NQR measurement to facilitate applied rf-field penetration.
However, powdering could damage the sample.
Thus, the present sample was crushed into coarse powder using a newly prepared sample.
The NQR spectrum of the new sample, that exhibits a full width at half maximum of $\sim65$ kHz, is three times narrower than for the previous sample, confirming the new samples better quality than before.
Here, the previous and present samples are labeled \#1 and \#2, respectively.
The SC transition temperature $T_{c}=1.85$ K are confirmed from the resistivity measurement to be the same for both samples.

\begin{figure}[tbh]
  \begin{center}
    \includegraphics[keepaspectratio=true,height=75mm]{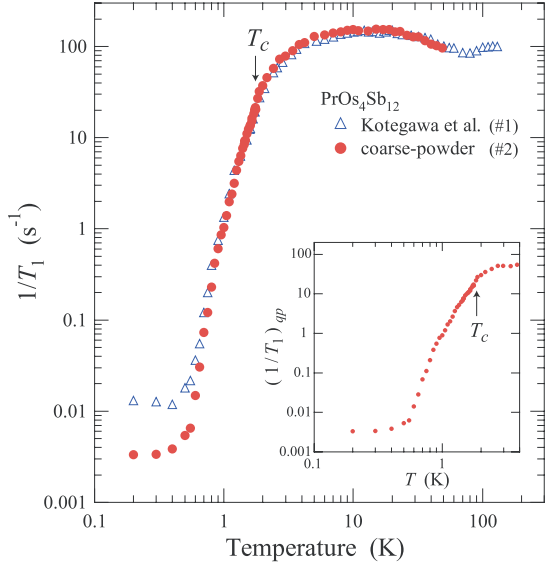}
  \end{center}
  \caption{$T$ dependences of $1/T_{1}$ for fine powder (open triangles) and coarse powder (closed circles) of PrOs$_{4}$Sb$_{12}$. Inset shows a plot of $(1/T_{1})_{qp}$ vs $T$. Here, $(1/T_{1})_{qp}$ is the quasi-particle contribution to $1/T_{1}$, which is evaluated from the relation $(1/T)_{qp} = (1/T_{1})_{obs} - A\exp (-\Delta_{\rm CEF}/T)$ (see text).\cite{Miyake}} 
\end{figure}

$T_1$ for $^{123}$Sb-$2\nu_{Q}$ transition ($\pm5/2 \leftrightarrow \pm3/2$) is uniquely determined from a theoretical curve for the recovery of the nuclear magnetization $m(t)$ in which the asymmetry parameters are incorporated.\cite{Chepin}
In the case of $I=7/2$ and $\eta=0.46$, the recovery curve is given by
\begin{eqnarray}
&&m_{2\nu_Q}(t)=\frac{M(\infty )-M(t)}{M(\infty )}\nonumber \\
&&= 0.077\exp\left(\frac{-3t}{T_{1}} \right)+0.0165\exp\left(\frac{-8.562t}{T_{1}} \right)\nonumber \\
&&+0.9065\exp \left(\frac{-17.207t}{T_{1}} \right).\label{Recov}
\end{eqnarray}
Figure 3 shows the $T$ dependences of $1/T_{1}$ for \#1 (open triangles) and \#2 (closed circles).
$1/T_{1}$'s for both samples exhibiting the same $T$ dependence, exponentially decreases in the range of $T = 0.7 \sim 1.85$ K with no coherence peak just below $T_{c}$.
Note that the present sample shows a more pronounced decrease in $1/T_{1}$ than the previous one; however, the saturation in $1/T_{1}$ was confirmed  below $T \sim 0.6$ K for both samples.
The $1/T_{1}=const.$ behavior at a very low $T$ used to be associated with local magnetic fluctuations induced by some magnetic impurities.
If this were the origin of $1/T_{1}=const.$ behavior, $1/T_{1}$ would be suppressed by applying a small magnetic field because such local magnetic fluctuations are depressed by the field.
However, the $1/T_{1}=const.$ in $T=0.2-0.5$ K is enhanced at $H \sim 0.1$ T, which indicates that it is of intrinsic origin.
In addition, note that a pronounced enhancement in $H_{c1}(T)$ was reported at $T \sim 0.6$ K by Cichorek \textit{et al.},\cite{Cichorek} and a small decrease in the penetration depth was also reported by Chia \textit{et al}.\cite{Chia}
These suggest an unknown intrinsic origin for the saturation of $1/T_{1}$ below $T \sim 0.6$ K.
\begin{figure}[tbh]
  \begin{center}
    \includegraphics[keepaspectratio=true,height=40mm]{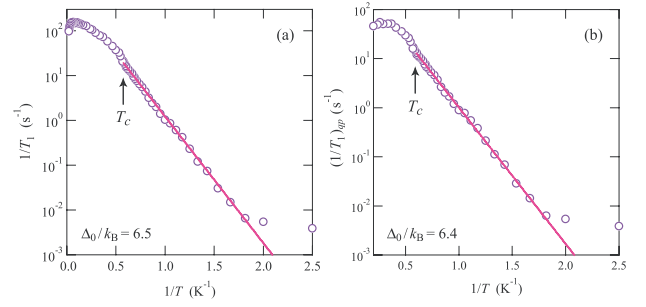}
  \end{center}
  \caption{Arrhenius plots of $1/T_{1}$ vs $1/T$ for PrOs$_{4}$Sb$_{12}$ \#2; (a) the raw data of $T_{1}$ and (b) the quasi-particle contribution $(T_1)_{qp}$ (see text). Solid lines indicate the relation $1/T_{1}\propto \exp (-\Delta_{0}/k_{\rm B}T)$ with (a) $\Delta_{0}/k_{\rm B} = 6.5$ K and (b) $\Delta_{0}/k_{\rm B} = 6.4$ K.}
\end{figure}

Figure 4(a) shows an Arrhenius plot of the raw data of $1/T_{1}$ vs $1/T$, demonstrating an exponential decrease following $1/T_{1}\propto \exp (-\Delta_{0}/k_{\rm B}T)$ down to 0.6 K below $T_{c}=1.85$ K.
Note that $1/T_{1}$ in this compound is determined by two contributions to relaxation that arise from quasi-particles being responsible for the onset of SC and low-lying CEF excitations as described by $(1/T_1)_{obs} = (1/T_1)_{qp} + A\exp (-\Delta_{\rm CEF}/T)$.\cite{Miyake}
The latter occurs because the CEF splitting between the singlet-$\Gamma_{1}$ ground state and the triplet-$\Gamma_{4}^{(2)}$ first-excited state is as small as $\Delta_{\rm CEF}\sim 8$ K.
Using $\Delta_{\rm CEF}=8$ K, $(1/T_{1})_{qp}$ is evaluated as shown in the inset of Fig. 3.
$(1/T_{1})_{qp}$ shows an exponential decrease as shown in Fig. 4(b).
The energy gap $2\Delta_{0}/k_{\rm B}T_{c}=6.9$ evaluated from the slope of the solid line reveals that PrOs$_{4}$Sb$_{12}$ is a strong-coupling superconductor.  
Although various experiments suggested the presence of point nodes, the exponential decrease in $(1/T_{1})_{qp}$ suggests that a full gap exists over a large part of the FS. 

\subsection{Evidence of multiband superconductivity (MBSC)}

Next, we focus on the evolution of SC characteristics caused by La substitution for Pr.
The recovery curve of nuclear-magnetization data for determining the $T_1$ of Sb(1) in the Pr-cage cannot be fitted by eq. (\ref{Recov}).
Therefore, short and long components of $T_{1}$, $T_{1}^{s}$ and $T_{1}^{l}$ are deduced from the superposition of two recovery functions described as,
\begin{eqnarray}
\frac{M(\infty )-M(t)}{M(\infty )}\nonumber = \sum_{i=s,l}a_{i} \biggr[ 0.077\exp\left(\frac{-3t}{T_{1}^{i}} \right) \nonumber\\
+0.0165\exp\left(\frac{-8.562t}{T_{1}^{i}} \right) + 0.9065\exp \left(\frac{-17.207t}{T_{1}^{i}} \right) \biggr],\label{Recov2}
\end{eqnarray}
where $a_{i}$ depends on NQR frequency and the La concentration $x$, but not on temperature.
\begin{figure}[tbh]
  \begin{center}
    \includegraphics[keepaspectratio=true,height=65mm]{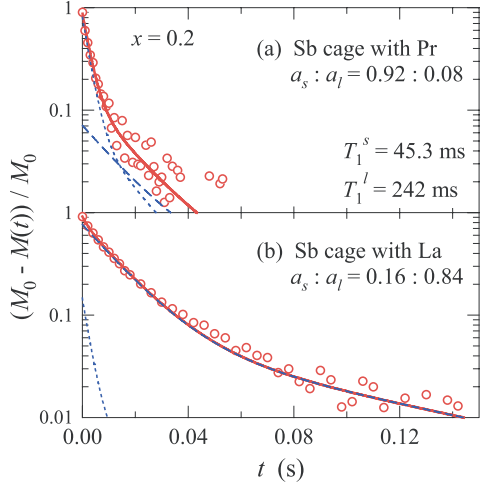}
  \end{center}
  \caption{$^{123}$Sb nuclear magnetization recovery curve measured at peaks of (a) Sb(1) and (b) Sb(2). The dotted and broken curves indicate the recovery curves due to the fast and slow relaxations, respectively. The solid curve indicates the sum of the two components.}
\end{figure}
Figures 5(a) and 5(b) show the nuclear magnetization recovery data for $x=0.2$ at 1.95 K for Sb(1) and Sb(2), respectively.
Because of a rapid decrease in $T_{1}^{s}$, the recovery data of Sb(2) can be fitted by almost single component with $T_{1}^{l}$ as shown in the broken line in Fig. 5(b).
Using the obtained $T_{1}^{l}$ and eq.(\ref{Recov2}), $T_{1}^{s}$ is determined from the recovery data of Sb(1).
Both data are well fitted by eq.(\ref{Recov2}) using the same values of $T_{1}^{l}$ and $T_{1}^{s}$ and different values of $a_{s}$ and $a_{l}$, as shown by the solid line in Fig. 5.
The respective errors of $T_{1}^{l}$ and $T_{1}^{s}$ are estimated to be about $1\sim 2$\% and $4\sim 5$\% by least-squares optimization.

\begin{figure}[tbh]
  \begin{center}
    \includegraphics[keepaspectratio=true,height=75mm]{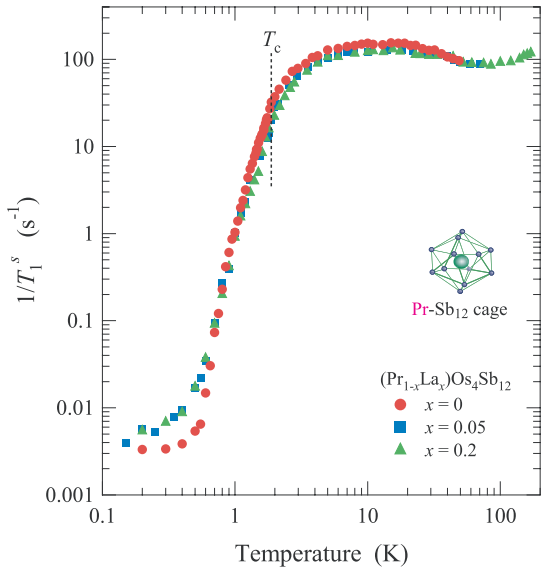}
  \end{center}
  \caption{$T$ dependences of $1/T_{1}^{s}$ for (Pr$_{1-x}$La$_{x}$)Os$_{4}$Sb$_{12}$; $x=0$ (circles), $x=0.05$ (squares) and $x=0.2$ (triangles).}
\end{figure}

Figure 6 shows the $T$ dependences of $1/T_{1}^{s}$ for $x = 0.05$ and 0.2.  
As expected, these $1/T_{1}^{s}$'s down to $\sim 0.7$ K ($\approx 0.4 T_{c}$) resemble the data of PrOs$_{4}$Sb$_{12}$, which reveal an exponential decrease in $1/T_{1}$ with the absence of a coherence peak just below $T_{c}$.
Therefore, hereafter, the $1/T_{1}^{s}$ of Sb(1) is denoted as $1/T_{1}^{Pr}$.
Note that a marked difference appears below $\sim 0.6$ K between $1/T_{1}^{Pr}$'s for $x=0.05$ and 0.2 and the $1/T_{1}$ for the pure compound.
The $1/T_{1}^{Pr}T=const.$ behavior for $x=0.05$ and 0.2 points to the appearance of residual density of states (RDOS) at the Fermi level induced by La substitution for Pr.
\begin{figure}[tbh]
  \begin{center}
    \includegraphics[keepaspectratio=true,height=75mm]{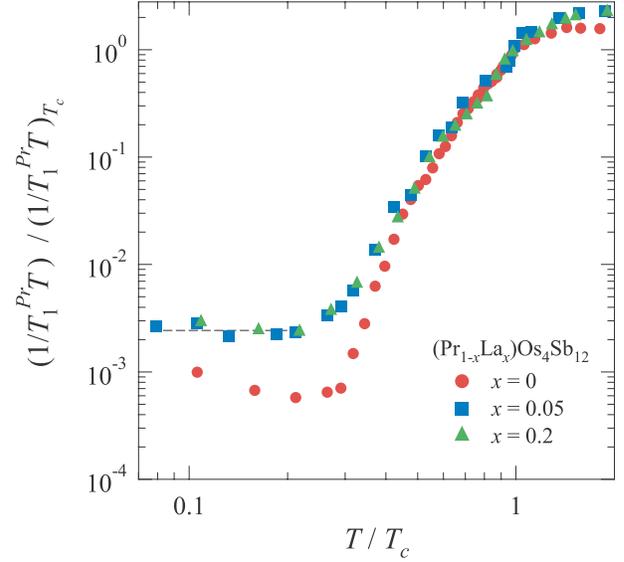}
  \end{center}
  \caption{Plots of $(1/T_{1}^{Pr}T)/(1/T_{1}^{Pr}T)_{T_c}$ vs $T/T_c$ for $x=0$ (circles), $x=0.05$ (squares) and $x=0.2$ (triangles).}
\end{figure}

Figure 7 shows the plots of  $(1/T_{1}^{Pr}T)/(1/T_{1}^{Pr}T)_{T_c}$ vs $T/T_c$ for $x=0$, 0.05 and 0.2.
The $1/T_{1}^{Pr}T=const.$ behavior is seen below $\sim 0.2T_{c}$ for $x=0.05$ and 0.2 with almost the same values.
For most HF superconductors with line nodes, it was reported that non magnetic impurities like La substitution for Ce induce RDOS in the SC state, and that RDOS increases as impurity content increases.\cite{Ishida_SrRuO}
Here, note that RDOS is estimated using
\begin{eqnarray}
\frac{N_{\rm res}}{N_{0}}=\sqrt{\frac{(T_{1}T)_{T_{c}}}{(T_{1}T)_{low-T}}},
\end{eqnarray}
where $N_{0}$ is the DOS in the normal state estimated from $(T_{1}T)$ at $T_{c}$, and $N_{\rm res}$ is the RDOS estimated from $(T_{1}T)_{low-T}=const.$ being valid well below $T_c$.

For the present compounds, however, RDOS is invariant even though La content increases from 5\% to 20\%, which demonstrates that the origin of RDOS is not the impurity effect characteristic for unconventional superconductivity that has been reported thus far.
Rather, the appearance of RDOS which is insensitive to La content suggests MBSC as revealed in several experiments.\cite{Seyfarth, Measson}
If we assume an MBSC model where a large SC gap exists over a large part of the FS, but an anisotropic gap with point nodes exists only in a small part, it is supposed that the latter gap is suppressed by either applying the magnetic field or substituting La for Pr.
Tentatively, the amount of the small part relative to the total of FS is estimated to be $\sim$ 5.5\% from $\sqrt{(T_{1}^{Pr}T)_{T_{c}}/(T_{1}^{Pr}T)_{low-T}}$. 

\begin{figure}[tbh]
  \begin{center}
    \includegraphics[keepaspectratio=true,height=75mm]{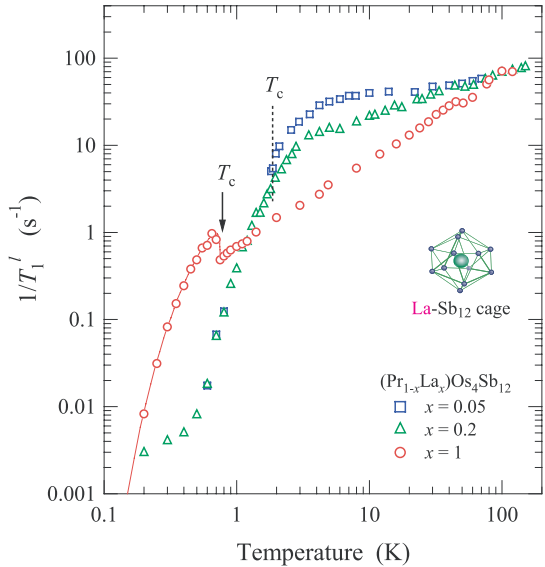}
  \end{center}
  \caption{$T$ dependences of $1/T_{1}^{l}$ for (Pr$_{1-x}$La$_{x}$)Os$_{4}$Sb$_{12}$; $x=0.05$ (squares), $x=0.2$ (triangles) and $x=1$ (circles). The solid line for $x=1$ indicates a calculation based on the  BCS model incorporating a gap anisotropy with $2\Delta_{0}/k_{\rm B}T_{c}=3.45$ and $(\Delta_{0}-\delta )/\Delta_{0}=0.30$ (see text).}
\end{figure}

In contrast to the result of Sb(1) in the Pr-cage, the $1/T_{1}^{l}$ for Sb(2) in the La-cage undergoes a remarkable concentration dependence as shown in Fig. 8.
Here, note that the data below $T_c$ at $x=0.05$ are not available because of the poor signal-to-noise ratio due to the Meissner shielding effect.
In the normal state, $1/T_{1}^{l}$'s at $x=0.05$ and 0.2 approach the value of LaOs$_{4}$Sb$_{12}$ above $\sim 80$ K, indicating that a $4f^{2}$-electron-derived contribution appears below $\sim 80$ K.
Therefore, hereafter, $1/T_{1}^{l}$ of Sb(2) is denoted as $1/T_{1}^{La}$.
In the SC state, the $T$ dependence of $1/T_{1}^{La}$ for Sb(2) resembles that for Sb(1) which reveals an exponential decrease with no coherence peak just below $T_{c}$, followed by a $1/T_{1}^{La}T=const.$ behavior below $\sim 0.6$ K.
The same $2\Delta/k_{\rm B}T_{c}=5.2$ is estimated for Sb(1) and Sb(2) for $x=0.2$ with $T_{c} = 1.80$ K. 

\subsection{Evolution of superconducting characteristics against La substitution}

\begin{fullfigure}[tbh]
  \begin{center}
    \includegraphics[keepaspectratio=true,height=75mm]{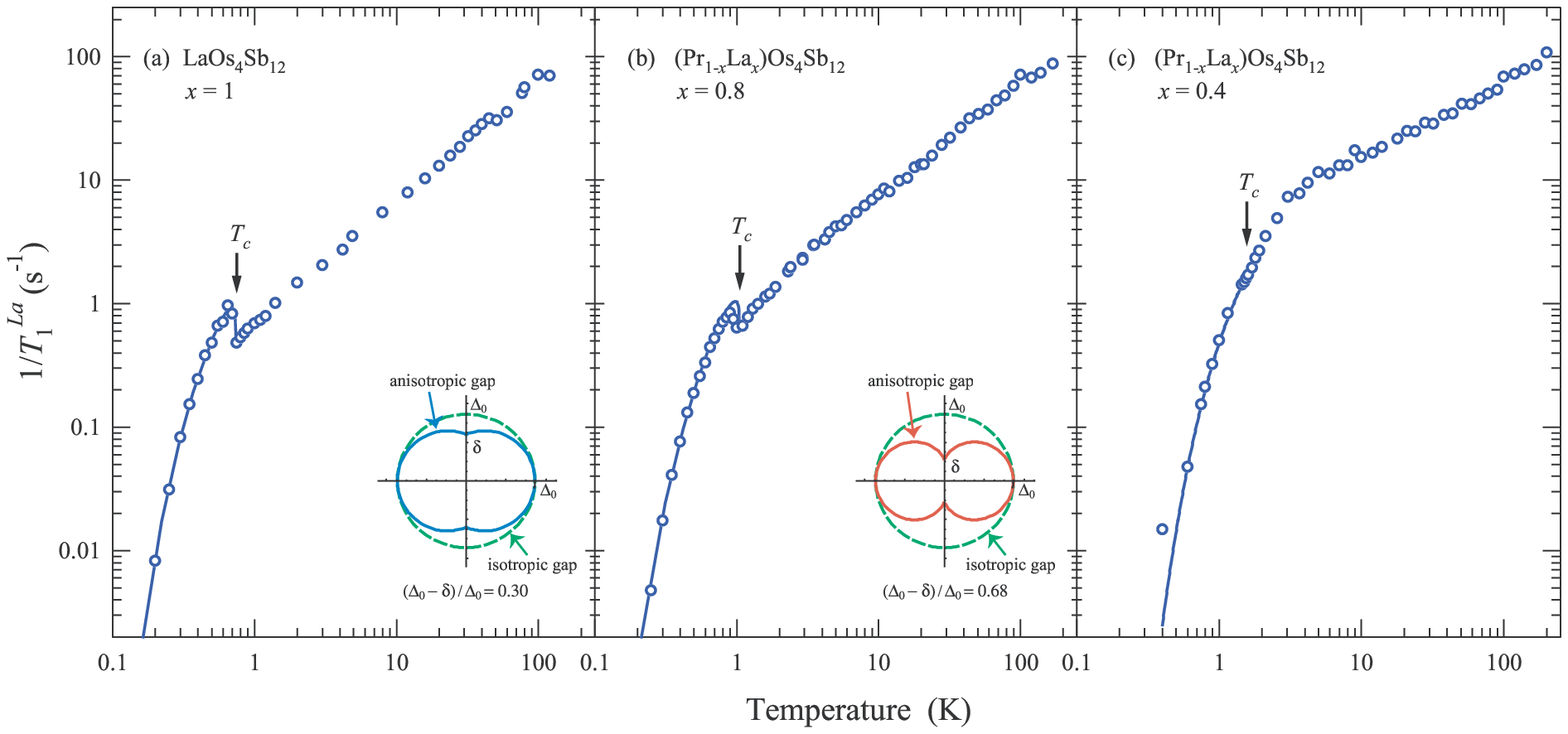}
  \end{center}
  \caption{$T$ dependences of $1/T_{1}^{La}$ for (Pr$_{1-x}$La$_{x}$)Os$_{4}$Sb$_{12}$; (a) $x=1$, (b) $x=0.8$ and (c) $x=0.4$. Solid lines in (a) and (b) indicate a calculation based on the BCS model incorporating a gap anisotropy: (a) $2\Delta_{0}/k_{\rm B}T_{c}=3.45$ and $(\Delta_{0}-\delta )/\Delta_{0}=0.30$, (b) $2\Delta_{0}/k_{\rm B}T_{c}=3.40$ and $(\Delta_{0}- \delta)/\Delta_{0}=0.68$. The solid line in Fig. 9(c) indicates a calculation using the relation $1/T_{1}^{La} \propto \exp(-\Delta_{0}/k_{\rm B}T)$ with $2\Delta_{0}/k_{\rm B}T_{c}=4.2$.}
\end{fullfigure}

Figure 9(a) shows the $T$ dependence of $1/T_{1}$ for LaOs$_{4}$Sb$_{12}$ ($x=1$), which obeys the Korringa relation, i.e., $1/T_{1}T=const.$ in the normal state and shows an exponential decrease with a coherence peak just below $T_{c}=0.74$ K, characteristic of conventional BCS $s$-wave superconductors.
The $1/T_1^{La}$ result for $x=0.8$ in Fig. 9(b) also shows a coherence peak just below $T_{c}\approx 1.05$ K, giving clear evidence of $s$-wave superconductivity in the case of the 20\% Pr compound.
However, the amplitude of the coherence peak becomes smaller than that of LaOs$_{4}$Sb$_{12}$.

In general, $1/T_{1}$ in the SC state is given by
\begin{eqnarray}
\frac{T_{1c}}{T_{1}}=\frac{2}{k_{\rm B}T_{c}}\int_{0}^{\infty}\left[ N_{\rm s}^{2}(E)+M_{\rm s}^{2}(E)\right]f(E)(1-f(E))dE
\end{eqnarray}
where $T_{1c}$ is $T_{1}$ at $T=T_{c}$, and $N_{s}(E)$ and $M_{s}(E)$ are respectively the quasi-particle DOS in the SC state and the anomalous DOS originating from the coherence effect of the transition probability characteristic for the singlet pairing state expressed as
\begin{equation}
N_{s}(E) = \frac{N(0)}{4\pi}\int_{0}^{2\pi}\int_{0}^{\pi}\frac{E}{\sqrt{E^2-\Delta^2(\theta, \phi)}}\sin\theta d \theta d \phi ,\label{Ns}
\end{equation} 
\begin{equation}
M_{s}(E) = \frac{N(0)}{4\pi}\int_{0}^{2\pi}\int_{0}^{\pi}\frac{\Delta (\theta, \phi)}{\sqrt{E^2-\Delta^2(\theta, \phi)}}\sin\theta d \theta d \phi , \label{Ms}
\end{equation}
For the $s$-wave with an isotropic gap $\Delta_{0}$, eqs. (\ref{Ns}) and (\ref{Ms}) diverge at $E=\Delta_{0}$, giving rise to a divergence of $1/T_{1}$ just below $T_{c}$.
However, this divergence of $1/T_{1}$ is suppressed by the life-time effect of quasi-particles, which is ascribed to the electron-phonon and/or electron-electron interactions, and the anisotropy of the energy gap due to the crystal structure broadens the quasi-particle DOS.

The $1/T_{1}^{La}$'s below $T_{c}$ for $x=0.8$ and 1 are fitted in terms of the BCS model with an anisotropy of energy gap amplitudes, as depicted in the insets of Figs. 9(a) and 9(b), respectively.
Since the presence of point-node gap is suggested for PrOs$_{4}$Sb$_{12}$, a point-node-like anisotropy is assumed for their gap functions expressed as
\begin{equation}
\Delta(\theta)=\delta+(\Delta_{0}-\delta)\sin\theta,
\end{equation} 
where $\Delta_{0}$ and $\delta$ are the maximum and minimum SC gaps, respectively.
If $\delta$ becomes zero, this gap function has the same shape as that in the $^{3}$He A-phase with two point-node gap.
This model explains well the data for $x=1$ and 0.8 as shown by the solid line in Figs. 9(a) and 9(b) which were calculated using $2\Delta_{0}/k_{\rm B}T_{c}$ = 3.45 and 3.40 with  $(\Delta_{0}-\delta)/\Delta_{0}=0.30$ and 0.68 for $x=1$ and 0.8, respectively.
With further increase in Pr content up to $x=0.4$, note that the coherence peak collapses as shown in Fig. 9(c), whereas energy gap is estimated as $2\Delta_{0} = 4.2k_{\rm B}T_{c}$ by fitting the data below $T_c$ to $1/T_{1}^{La} \propto \exp(-\Delta_{0}/k_{\rm B}T)$ that is shown by the solid line in Fig. 9(c). 
These results reveal that the increase in Pr content suppresses the appearance of coherence peak due to the increase in $2\Delta_{0}/k_{\rm B}T_{c}$ and at the same time yields an anisotropy of the SC gap.
This is because the Pr substitution for La makes a coupling-forming Cooper pair strong, leading to a strong-coupling superconductivity in PrOs$_4$Sb$_{12}$.  

\begin{fulltable}[tb]
\caption{ Physical properties of (Pr$_{1-x}$La$_{x}$)Os$_{4}$Sb$_{12}$ compounds. $x$ is the concentration of La. $T_{c}$ is an SC transition evaluated from the midpoint of the resistivity decrease. Sb(1) and Sb(2) are the respective sites forming  Pr and La cages.  $2\Delta_{0}/k_{\rm B}T_{c}$ is evaluated using the anisotropic $s$-wave model for $x=0.8$ and 1 (see the text), and using the relation $1/T_{1} \propto \exp (-\Delta_{0}/k_{\rm B}T)$ for $x\le 0.6$.  $N_{res}$ is the RDOS evaluated from the expression $N_{\rm res}/N_{0}=\sqrt{(T_{1}^{Pr}T)_{T_{c}}/(T_{1}^{Pr}T)_{low-T}}$ (see text).  }
\label{t3}
\begin{tabular}{cccccccc}
\hline
 &  & PrOs$_{4}$Sb$_{12}$ &  &  &  &  & LaOs$_{4}$Sb$_{12}$ \\
\cline{3-8}
\multicolumn{2}{c}{$x$} & 0 & 0.05 & 0.2 & 0.4 & 0.8 & 1 \\
\hline
\multicolumn{2}{c}{$T_{c}$ (K)} & 1.85 & 1.84 & 1.80 & $\sim$1.67 & $\sim$1.05 & 0.74 \\
\raisebox{-1.8ex}[0pt][0pt]{Sb(1)} & $2\Delta_{0} /k_{\rm B}T_{c}$ & 6.9 & 5.9 & 5.2 & - & - &  \\
 & $N_{\rm res}$ (\%) & $< 2.7$ & $5\sim 6$ & $5\sim 6$ & - & - &  \\
\raisebox{-1.8ex}[0pt][0pt]{Sb(2)} & $2\Delta_{0} /k_{\rm B}T_{c}$ &  & - & 5.2 & $\sim$4.2 & 3.40 & 3.45 \\
 & $N_{\rm res}$ (\%) &  & - & $7\sim 8$ & - & - & 0 \\
\hline
\end{tabular}
\end{fulltable}

Table \ref{t3} shows a summary of the SC properties of (Pr$_{1-x}$La$_{x}$)Os$_{4}$Sb$_{12}$.
Here, $2\Delta_{0}/k_{\rm B}T_{c}$ is estimated using the anisotropic $s$-wave model for $x=0.8$ and 1, and the relation $1/T_{1} \propto \exp (-\Delta_{0}/k_{\rm B}T)$ for $x\le 0.6$.
Because of the continuous decrease in $1/T_{1}$ above $T_{c}$, the determination of $T_{c}$ from $1/T_{1}$ is difficult for $x\le 0.6$, therefore,  $T_{c}$ was determined from the midpoint of the resistivity decrease.

\section{Discussions}

\subsection{Normal state properties}

\begin{figure}[tbh]
  \begin{center}
    \includegraphics[keepaspectratio=true,height=100mm]{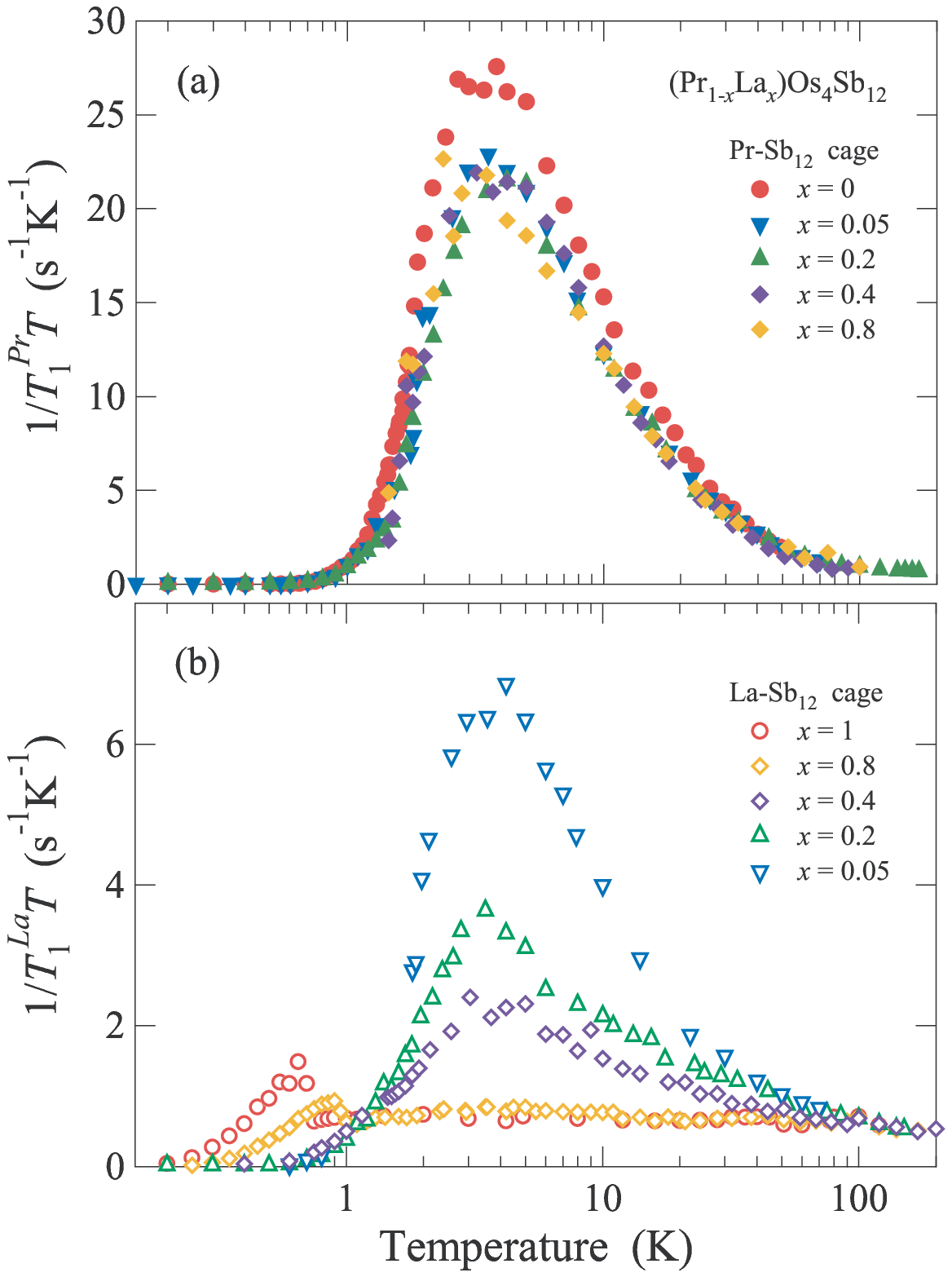}
  \end{center}
  \caption{$T$ dependences of $1/T_{1}T$ for (Pr$_{1-x}$La$_{x}$)Os$_{4}$Sb$_{12}$ ($x=0$, 0.05, 0.2, 0.4, 0.8, 1); (a) $1/T_{1}^{Pr}T$ at Sb(1) and (b) $1/T_{1}^{La}T$ at Sb(2).}
\end{figure}

To observe the evolution of normal and SC properties caused by La substitution for Pr, $1/T_{1}T$'s for all contents are plotted in Figs. 10(a) and 10(b).
For Sb(1) in the Pr cage, $1/T_{1}^{Pr}T$'s develop upon cooling below $\sim 100$ K.
$1/T_{1}^{Pr}T$'s in (Pr$_{1-x}$La$_{x}$)Os$_{4}$Sb$_{12}$ show similar $T$ dependences with a maximum at approximately 3 K regardless of La content, although the value for the peak is slightly smaller than the value for pure PrOs$_{4}$Sb$_{12}$.
These results reveal that Sb(1) is predominated by $4f^2$-derived magnetic fluctuations that are rather localized in the Pr cage.
In contrast, $1/T_{1}^{La}T$'s for Sb(2) in the La cage that are shown in Fig. 10(b) are markedly suppressed as La content increases.
Note that $1/T_1^{La}T$ for $x=0.05$ exhibits a maximum at around 3 K as well as that at Sb(1).
This means that Sb(2) in Pr-rich compounds is predominated by $4f^2$-derived magnetic fluctuations, whereas Sb(2) in La-rich compounds is not affected by  $4f^2$-derived magnetic fluctuations even though Pr content increases up to 0.2.  
These contrasting behaviors of $1/T_{1}T$ are because $4f^{2}$-derived magnetic fluctuations are almost localized in association with the low-lying CEF state with a quadrupole degree of freedom, which might be important for the unconventional SC state of PrOs$_{4}$Sb$_{12}$.

\subsection{Novel superconducting characteristics}

One of the unconventional SC properties of PrOs$_{4}$Sb$_{12}$ is the absence of a coherence peak in $1/T_1$ just below $T_{c}$.\cite{Kotegawa}
It is, however, known that there are several reasons for this.
For example, the coherence peak in the anisotropic $s$-wave superconductor CeRu$_{2}$ is markedly suppressed because of the anisotropy of SC gap.\cite{Mukuda}
When this anisotropic SC gap is smeared out by substituting nonmagnetic impurities of Al for Ru, the coherence peak is enhanced.
Furthermore, in the case of strong-coupling $s$-wave superconductors, the appearance of a coherence peak is suppressed due to the life time effect of quasi-particles through the electron-phonon damping channel.
$1/T_{1}$ of TlMo$_{6}$Se$_{7.5}$ shows an exponential decrease over four decades below $0.8T_{c}$ with $2\Delta_{0}=4.5k_{\rm B}T_{c}$, but does not exhibit the coherence peak just below $T_{c}$.\cite{Ohsugi}
From the large $2\Delta_{0}=6.9k_{\rm B}T_{c}$, since PrOs$_{4}$Sb$_{12}$ is classified as a strong-coupling superconductor, it is possible that the coherence peak is suppressed. 
Moreover, the fact that $T_{c}$ is almost invariant against substituting La for Pr suggests that an $s$-wave-like gap is predominant in PrOs$_{4}$Sb$_{12}$.  

\begin{figure}[tbh]
  \begin{center}
    \includegraphics[keepaspectratio=true,height=70mm]{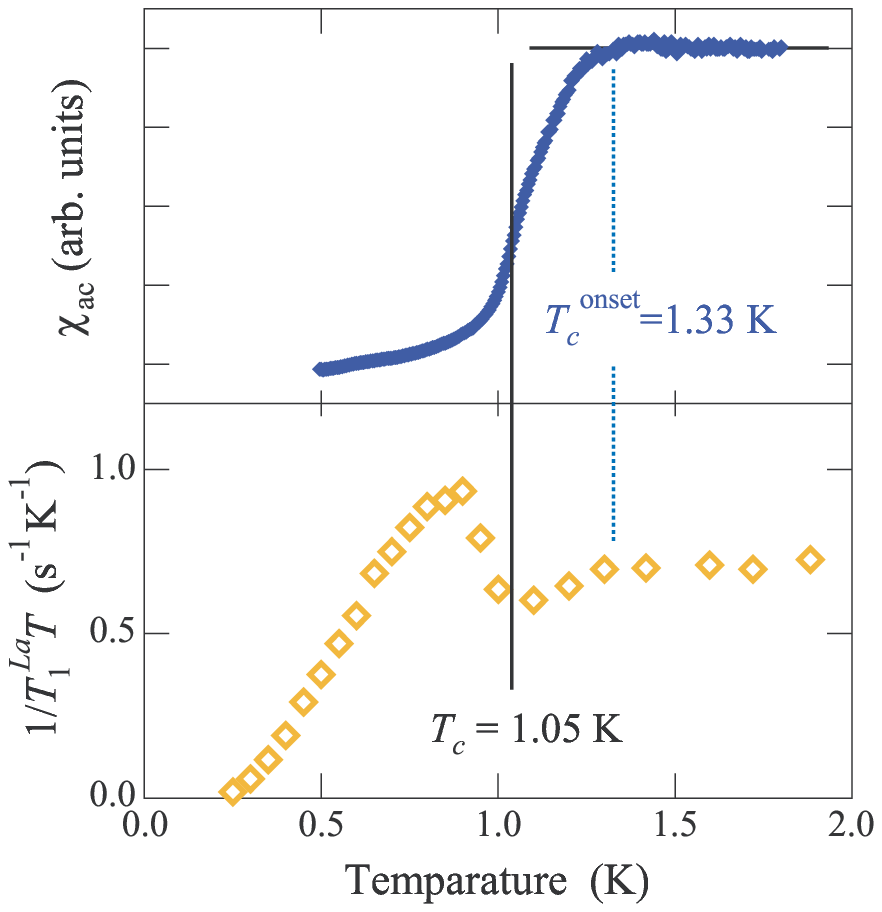}
  \end{center}
  \caption{$T$ dependences of $\chi_{ac}$ and $1/T_{1}^{La}T$ for $x=0.8$. $\chi_{ac}$ was measured using an NQR coil at a resonance frequency of $\sim 60$ MHz.}
\end{figure}

From other context, we present the evolution from weak-coupling superconductivity to strong-coupling superconductivity for the Pr substitution for LaOs$_{4}$Sb$_{12}$. 
Figure 11 shows the $T$ dependences of $1/T_{1}^{La}T$ and the ac susceptibility $\chi_{ac}$ for the 20\% Pr compound.
The pronounced coherence peak in $1/T_{1}^{La}T$ at $T_{c}\approx 1.05$ K indicates that the $s$-wave superconductivity set in.
Since $T_{c}$ for this sample ($x=0.8$) is 1.4 times larger than $T_c=0.74$ K for LaOs$_{4}$Sb$_{12}$, the Pr substitution increases $T_c$.
Furthermore, note that $T_{c}^{onset}=1.33$ K below which the SC diamagnetism that appears is significantly larger than $T_{c}\approx 1.05$ K, determined by Sb(2)-$T_1$ measurement.
Since the enhancement in $T_c$ is due to the Pr substitution, one reason the onset of superconductivity takes place at a temperature higher than the bulk $T_c$ is that the Pr cage enhances an SC pairing interaction via the local interaction between Pr ions and conduction electrons. 

Next we consider the possibility of MBSC in PrOs$_{4}$Sb$_{12}$.
The $1/T_{1}^{Pr}T=const.$ behavior observed for $x=0.05$ and 0.2 well below $T_{c}$ indicates the presence of RDOS, which is insensitive to La content, as shown in Fig. 7.
In addition, the exponential decrease in $1/T_{1}^{Pr}$ below $T_{c}$ is robust against La substitution for Pr.
These results are consistent with a nodeless SC gap on the large part of FS which exists in both PrOs$_{4}$Sb$_{12}$ and LaOs$_{4}$Sb$_{12}$.
The anisotropic SC gap with point nodes is relevant with the small part of FS inherent in PrOs$_{4}$Sb$_{12}$, which amounts to $\sim$ 5.5\% of the total DOS.
Since this small anisotropic gap collapses by either applying a magnetic field or substituting La for Pr, the small FS inherent in PrOs$_{4}$Sb$_{12}$
does not play any primary role because $T_c$ does not appreciably decrease by La substitution for Pr.
From these results, it is apparent that the HF superconductivity of PrOs$_{4}$Sb$_{12}$ differs from the previous examples where a $4f$-electron-derived HF band plays an important role.
Instead, the MBSC model is rather promising for interpreting the novel SC properties of (Pr$_{1-x}$La$_{x}$)Os$_{4}$Sb$_{12}$.

\section{Conclusions}

We have reported on the systematic evolution of normal-state properties and SC characteristics in the filled-skutterudite compounds (Pr$_{1-x}$La$_{x}$)Os$_{4}$Sb$_{12}$, determined using Sb nuclear-quadrupole-resonance (NQR) experiments.
The temperature dependence of $1/T_{1}$ of Sb nuclei was separately measured for Pr and La cages.
In the normal state for Pr-rich compounds of $x=0.05$ and 0.2, the $T$ dependence of $1/T^{Pr}_1T$ for Pr cage has revealed almost the same behavior as that for pure PrOs$_{4}$Sb$_{12}$ regardless of the increase in La content.
In contrast, the $1/T^{La}_1T$ for La cage is strongly suppressed with increasing La concentration.
These results show that $4f^{2}$-derived magnetic fluctuations are almost localized at the Pr site.
In their SC state, $1/T_{1}^{Pr}$ exponentially decreases down to $T=0.7$ K with no coherence peak below $T_{c}$ as well as in PrOs$_{4}$Sb$_{12}$.
The remarkable finding is that RDOS at the Fermi level below $T_c$ is induced by La substitution for Pr.
From the fact that the amount of RDOS does not increase and $T_c$ does not decrease with an increase in La content, it is concluded that the RDOS induced by La substitution is not due to the impurity effect used to be observed in unconventional superconductors with the line-node gap.
Rather, a small part of the FS which contributes to $\sim$ 5.5\% of the total DOS is suggested to become gapless for $x=0.05$ and 0.2, yielding RDOS.
These results are understood in terms of the multiband superconductivity (MBSC) model that has been proposed recently from the thermal-transport measurement under a magnetic field.
For La-rich compounds of $x=0.8$ and 1, on the other hand, the $1/T^{La}_1$ results exhibit a coherence peak and the nodeless energy gap characteristic for weak-coupling BCS $s$-wave superconductors.
With increasing Pr content, $T_c$ increases and the energy gap increases from $2\Delta_{0}/k_{\rm B}T_{c}= 3.45$ for the pure La compound to $2\Delta_{0}/k_{\rm B}T_{c}=4.2$ and 5.2 for the 60\% Pr and 80\% Pr compounds, respectively.
The Pr substitution for La enhances the pairing interaction and introduces an anisotropy into the energy-gap structure.
These results are proposed to be understood in terms of the MBSC model which assumes a full gap for the large part of FS and the presence of point nodes for the small $4f^2$-derived FS inherent in PrOs$_{4}$Sb$_{12}$.
The FS existing in PrOs$_{4}$Sb$_{12}$ could be successively connected to a part of the FS in LaOs$_{4}$Sb$_{12}$, and the anisotropic gap with the point nodes inherent in PrOs$_{4}$Sb$_{12}$ is markedly suppressed by either applying the magnetic field or substituting La for Pr.
The novel strong-coupling superconductivity in PrOs$_{4}$Sb$_{12}$ is suggested to be mediated by the local interaction between $4f^{2}$ low-lying CEF states with the electric quadrupole degree of freedom and conduction electrons.
This coupling causes a mass enhancement of quasi-particles for the large part of FS and induces the small FS with point nodes in the SC gap function of PrOs$_{4}$Sb$_{12}$.
However, this small FS does not play any primary role for the strong-coupling SC in PrOs$_{4}$Sb$_{12}$ because the collapse of this anisotropic gap due to the substitution of La for Pr does not decrease $T_c$ appreciably.
In this context, it is concluded that PrOs$_{4}$Sb$_{12}$ is the multiband superconductor with a well-developed gap on the large part of FS and the anisotropic gap on the small FS.

\section*{Acknowledgments}
We wish to thank G.-q. Zheng, H. Kotegawa, H. Tou, K. Ishida, K. Miyake and M. Nishiyama for their helpful and valuable comments and discussions.
We also thank H. Sumitani for measuring resistivity.
This work was partially supported by a Grant-in-Aid for Creative Scientific Research (15GS0213), MEXT, The 21st Century COE Program supported by the Japan Society for the Promotion of Science, and a Grant-in-Aid for Scientific Research Priority Area "Skutterudite" (Nos. 15072204 and 15072206), MEXT.

\end{document}